\documentclass[sigconf, nonacm]{acmart}
\usepackage[ruled,vlined]{algorithm2e}
\usepackage{lscape}
\usepackage{makecell}  %

\usepackage{subcaption}
\usepackage{longtable}

\usepackage[most]{tcolorbox}
\usepackage{xcolor}
\usepackage{longtable}
\usepackage{courier}
\usepackage{makecell}
\AtBeginDocument{%
  }

\acmConference[CUI '25]{Make sure to enter the correct
  conference title from your rights confirmation emai}{July 08--10,
  2025}{Waterloo, Canada}
\begin{document}

\title{Learn, Explore and Reflect by Chatting: \\Understanding the Value of an LLM-Based Voting Advice Application Chatbot}
\thanks{Preprint. This paper has been accepted to ACM Conversational User Interfaces
2025 (CUI '25); please cite that version instead.}

\author{Jianlong Zhu}
\affiliation{%
  \department{Department of Computer Science}
  \institution{Saarland University}
  \city{Saarbrücken}
  \country{Germany}
}
\email{jzhu@cs.uni-saarland.de}
\orcid{0000-0001-6109-8097}

\author{Manon Kempermann}
\affiliation{%
  \department{Department of Computer Science}
  \institution{Saarland University}
  \city{Saarbrücken}
  \country{Germany}
}
\email{make00009@stud.uni-saarland.de}

\author{Vikram Kamath Cannanure}
\affiliation{%
  \department{Department of Computer Science}
  \institution{Saarland University}
  \city{Saarbrücken}
  \country{Germany}
}
\email{cannanure@cs.uni-saarland.de}

\author{Alexander Hartland}
\affiliation{%
  \department{Department of European Social Research}
  \institution{Saarland University}
  \city{Saarbrücken}
  \country{Germany}
}
\email{alexander.hartland@uni-saarland.de}

\author{Rosa M. Navarrete}
\affiliation{%
  \department{Department of European Social Research}
  \institution{Saarland University}
  \city{Saarbrücken}
  \country{Germany}
}
\email{rosa.navarrete@uni-saarland.de}

\author{Giuseppe Carteny}
\affiliation{%
  \department{Department of European Social Research}
  \institution{Saarland University}
  \city{Saarbrücken}
  \country{Germany}
}
\email{giuseppe.carteny@uni-saarland.de}

\author{Daniela Braun}
\affiliation{%
  \department{Department of European Social Research}
  \institution{Saarland University}
  \city{Saarbrücken}
  \country{Germany}
}
\email{d.braun@uni-saarland.de}

\author{Ingmar Weber}
\affiliation{%
  \department{Department of Computer Science}
  \institution{Saarland University}
  \city{Saarbrücken}
  \country{Germany}
}
\email{iweber@cs.uni-saarland.de}

\renewcommand{\shortauthors}{Zhu et al.}

\begin{abstract}

Voting advice applications (VAAs), which have become increasingly prominent in European elections, are seen as a successful tool for boosting electorates’ political knowledge and engagement. However, VAAs’ complex language and rigid presentation constrain their utility to less-sophisticated voters. While previous work enhanced VAAs’ click-based interaction with scripted explanations, a conversational chatbot’s potential for tailored discussion and deliberate political decision-making remains untapped. Our exploratory mixed-method study investigates how LLM-based chatbots can support voting preparation. We deployed a VAA chatbot to 331 users before Germany’s 2024 European Parliament election, gathering insights from surveys, conversation logs, and 10 follow-up interviews. Participants found the VAA chatbot intuitive and informative, citing its simple language and flexible interaction. We further uncovered VAA chatbots’ role as a catalyst for reflection and rationalization. Expanding on participants’ desire for transparency, we provide design recommendations for building interactive and trustworthy VAA chatbots.

\end{abstract}


\maketitle

\section{Introduction}
Voters’ knowledge is the cornerstone of democracy. Since the 1980s, voting advice applications (VAAs) have educated users about parties' policy preferences and emerged as a successful digital civic education tool across Europe \cite{garzia_voting_2019, tromborg_candidates_2023}. Initially devised as a paper-and-pencil questionnaire, VAAs today offer a web interface to guide prospective voters to compare their policy preferences with those of parties of interest through a turn-by-turn opinion survey, and produce a ranked list or graphical representation of policy agreement \cite{garzia_voting_2019}. Digital VAAs are increasingly popular among voters in Germany, with \textit{Wahl-O-Mat}, created by the Federal Agency for Civic Education, being used 15.7 million times ahead of the 2021 federal election\footnote{https://www.bpb.de/die-bpb/presse/pressemitteilungen/340514/nutzungsrekord-beim-wahl-o-mat/} and 14.8 million times in the run-up to the European Parliament election in 2024\footnote{https://www.bpb.de/die-bpb/presse/pressemitteilungen/549326/wahl-o-mat-zur-europawahl-endet-mit-nutzungsrekord/}. Such digital interactions with VAAs have become integral to informing and educating prospective voters before an election.

VAAs are a convenient complement to comprehensive civic education programs \cite{fossen_whats_2014} as they conveniently engage and inform voters by comparing political parties based on policy positions \cite{ferreira_da_silva_three_2023}. Studies show that VAAs can enhance political knowledge \cite{munzert_meta-analysis_2021, waldvogel_what_2023}, boost overall political interest, and increase voting intention \cite{waldvogel_what_2023, stadelmann-steffen_role_2023}. However, VAAs tend to use complex terminology \cite{kamoen_i_2017}, cater to voters who are more educated and politically interested \citep{van_de_pol_beyond_2014, garzia_voting_2019, schultze_effects_2014}, and focus on a narrow set of policy issues \citep{wagner_matching_2012}. Their intended functionality also risks oversimplifying democratic processes by reducing complex decision-making to a quick, mouse-clicking exercise at odds with certain views of democracy \citep{fossen_whats_2014}.

To address some of these shortcomings, a scripted chatbot add-on to VAAs’ click-based interface has been explored as a way to help voters understand the underlying political context and boost political knowledge \cite{kamoen_i_2022, van_zanten_voting_2024}. Nonetheless, such an interface places a chatbot in an auxiliary role that does not allow for in-depth discussions. The possibility of large language models (LLMs) as a discussion tool that fosters reflection \cite{tanprasert_debate_2024,hadfi_augmented_2022, chiang_enhancing_2024, reicherts_extending_2022} brings the potential of enhancing the effectiveness and expanding the functionality of VAAs with LLMs, which calls for empirical research on AI assistance in such high-stake decision-making scenario. Along with the capabilities, however, LLMs are also prone to generating non-factual information (“hallucination”) on election-related facts \cite{helming_generative_2023, romano_dataset_2024, angwin_seeking_2024, anthropic_testing_2024}, and have been found to exhibit political bias \cite{feng_pretraining_2023,hartmann_political_2023, motoki_more_2023, chalkidis_llama_2024,rettenberger_assessing_2024}. Thus, it is crucial to understand how prospective voters may perceive and use an LLM-based VAA chatbot in a realistic setting to address these shortcomings.

Therefore, we formulated the following research questions: 
\begin{description}

\item[RQ1:] How can an LLM-based chatbot address known challenges to utilizing VAAs?

\item[RQ2:] What new opportunities can the conversational capabilities of LLMs bring to voting preparation?

\item[RQ3:] What are the obstacles to the trusted utilization of an LLM-based VAA chatbot?
\end{description}

\begin{figure*}[htbp]

    \begin{subfigure}[b]{0.7\textwidth}
        \centering
        \includegraphics[width=1\textwidth]{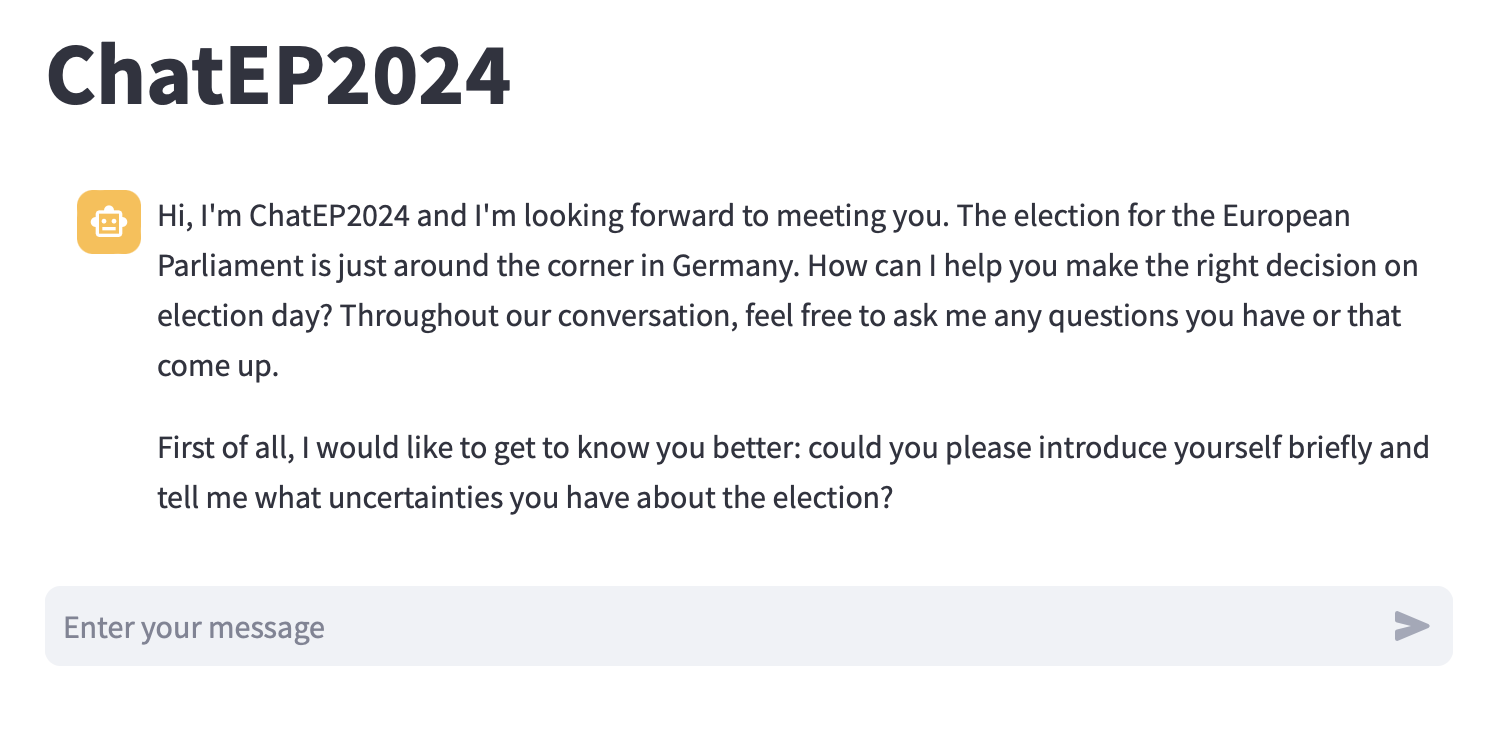}
        \caption{Initial screen and the start of the unstructured exchange.}
        \label{fig:screenshot_start}
    \end{subfigure}

    \vspace{0.5cm}

    \begin{subfigure}[b]{0.7\textwidth}
        \includegraphics[width=1\textwidth]{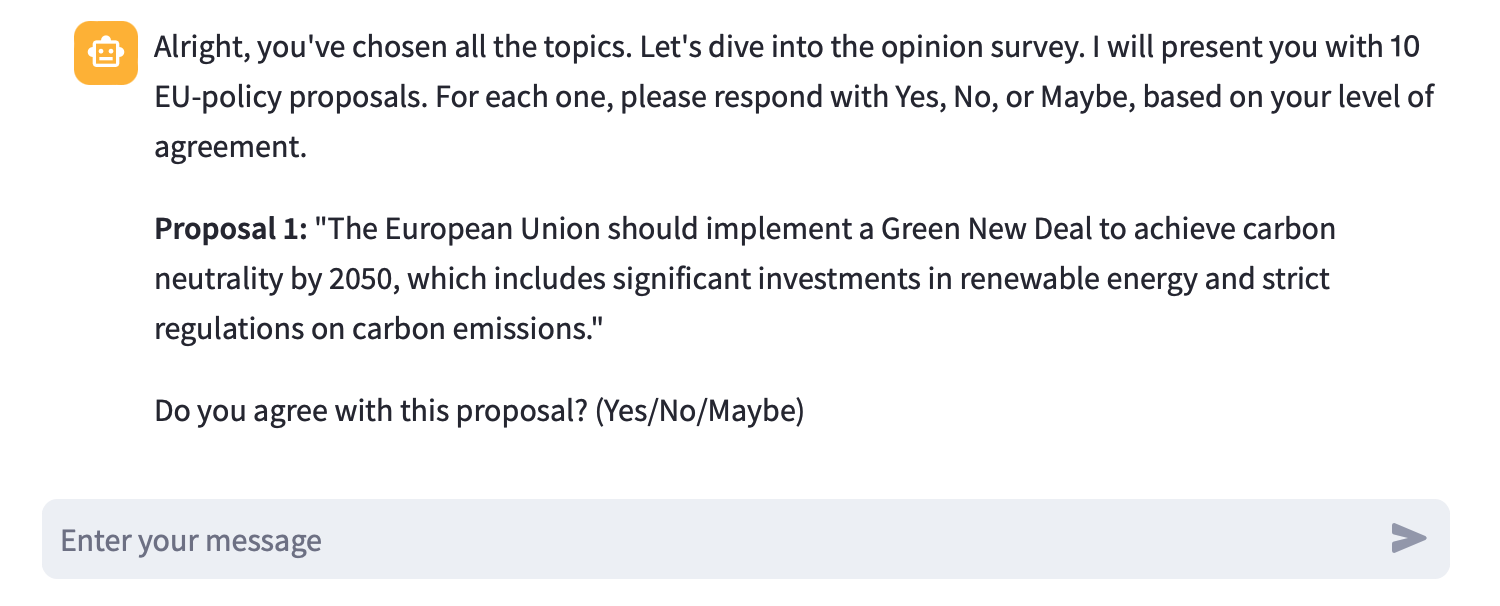}
        \caption{Start of the structured exchange.}
     \label{fig:screenshot_structured}

    \end{subfigure}
    \caption{Examples of guidance from the chatbot in English. The actual user study was performed in German.}
    \Description[Two screenshots from the prototype chatbot.]{The first screenshot shows the chatbot displaying a message starting with "I'm looking forward to meeting you" and ending with "please introduce yourself briefly and tell me what uncertainties you have about the election". In the second image, the chatbot presents a policy proposal and asks for the user's opinion.}
    
    \label{fig:screenshots}  
\end{figure*}

To address these questions, we employed a two-phase mixed-method study design. First, we recruited 331 participants in Germany to interact with our chatbot (Figure \ref{fig:screenshots}) and complete a survey on demographics and user experience. Then, we conducted phone interviews with 10 participants to explore voting preparation strategies, chatbot experiences, and perceptions of informedness and trust. Participants valued the chatbot's ability to provide \textbf{concise, clear overviews} on various topics, which enhanced their understanding of the political landscape. Participants with lower educational attainment and those with moderately low political self-efficacy were more likely to feel informed. The chatbot sparked \textbf{curiosity}, encouraged \textbf{reflection} and \textbf{rationalization} without necessarily altering opinions. Many pointed to LLMs’ overall lack of \textbf{truthfulness} and the chatbot’s lack of \textbf{disclosure} about sources and training as reasons to withhold trust. Most participants nevertheless indicated a willingness to use a similar chatbot again despite concerns.

Our contributions to the HCI and CUI communities are as follows:
\begin{itemize}
    \item We expand existing literature on voters' challenges with VAAs and provide an empirical basis for an LLM-based chatbot's effectiveness in enhancing political knowledge. Participants preferred a chatbot's comprehensible and personalized explanations, unlike traditional VAAs' overwhelming experience that deters voters from deep engagement. 
    
    \item Beyond the guided information seeking that can be expected from a web-based education tool, our deployment demonstrated the chatbot interface’s promise in stimulating active cognitive engagement, characterized by curiosity-driven exploration and reflection.
    
    \item We provide design recommendations to facilitate informative, deliberative and trusted interaction with LLM-based VAAs.
\end{itemize}

Our work provides empirical insights to help civic educators design a discussion chatbot that addresses voters’ information needs, stimulates cognitive engagement in learning, and meets requirements for trust. It also assists stakeholders in evaluating the competency and preparedness of the general public to utilize the technology with an awareness of its risks.

\section{Related Work}

\subsection{Functionality of Voting Advice Applications (VAAs)}

In European elections, the political landscape is typically complex, with many parties competing for representation and multi-party coalition governments often forming. Voters must navigate various policy options and party platforms to make informed decisions, contrasting with the more ideological two-party contests in countries such as the US \cite{riera_electoral_2022}. Voting advice applications (VAAs), which allow users to compare their policy preferences with political parties through an opinion survey, have become an increasingly popular civic education tool across Europe. VAAs serve as a convenient complement to classroom-style "wholesale programs in voter education" \cite{fossen_whats_2014} to allow prospective voters to quickly rank parties based on policy preferences without having to perform an extensive information search \cite{ferreira_da_silva_three_2023, tromborg_candidates_2023}. Previous studies show VAA usage increases political knowledge \cite{munzert_meta-analysis_2021, waldvogel_what_2023}, strengthens political interest \cite{waldvogel_what_2023}, and boosts voter turnout \cite{garzia_voting_2019, stadelmann-steffen_role_2023, munzert_meta-analysis_2021}.

\subsubsection{Limitations of Traditional VAAs}

Despite their popularity, traditional VAAs face challenges in achieving their intended goals due to limitations in terms of accessibility and effectiveness, which constrain their potential to foster informed democratic participation. First, users sometimes report comprehension problems with VAAs \cite{kamoen_i_2017}, as the phrasing of VAA statements assumes a certain level of political knowledge and familiarity with relevant political terminology. Second, VAAs are not evenly adopted across demographics, with users typically being young, educated, and politically engaged \cite{van_de_pol_beyond_2014, garzia_voting_2019, schultze_effects_2014, walder_explaining_2024}. The uneven uptake can be partly explained by the cognitive cost of interacting with VAAs, which is said to be reduced by higher education or political knowledge \cite{walder_explaining_2024}. Similarly, the effects of VAA use may be mediated by demographic factors, albeit with mixed evidence \cite{van_de_pol_beyond_2014, germann_voting_2023, munzert_online_2021}. Third, VAAs often rigidly present official statements or expert opinions on a select range of topics without tailoring to users' interests and information needs. The carefully worded policy aims may fail to clarify nuanced ideological differences or help assess non-policy attributes such as candidates' trustworthiness \cite{wagner_matching_2012}.

\subsection{LLM-Based Chatbot for Political Engagement: Promises and Gaps}

In contrast to the WIMP paradigm of user interface design that prizes simplicity, a conversational interface may be more aligned with the desired cognitive processes for active participation in high-stakes decision making \cite{reicherts_make_2020}. The click-based interfaces of traditional VAAs are designed to streamline the voting process by matching voters to pre-defined policy options. The VAA interfaces typically ask users to respond to each statement with a mouse click, optimizing speedy choices on polarized issues. Such a process, reflecting the ideal of a social choice model of democracy \cite{fossen_whats_2014}, can be seen as overly simplistic by supporters of more deliberative forms of democracy, who argue that democracy should encourage continuous discussion and refinement of opinions rather than merely aggregating existing preferences \cite{mutz_hearing_2006, fossen_whats_2014}. There have been calls to envision other forms of civic education tools that produce citizens who are "co-legislators" of a democracy, as opposed to "policy shoppers" \cite{fossen_whats_2014}, and a fully conversational VAA may provide inspiration.

User studies have shown that a chatbot displaying manually prepared answers that serves as an explainer to difficult concepts in a VAA can enhance political knowledge, especially for users with lower political sophistication \cite{kamoen_i_2022, van_zanten_voting_2024,hankel_hi_2024}. The proposed explainer chatbot is nevertheless not fully conversational because it serves only as a supplementary tool to assist users in navigating a click-based interface, rather than facilitating an open discussion. With the rise of LLMs, recent work has positioned LLMs as a tool for facilitating critical thinking, deliberation and decision-making \cite{tanprasert_debate_2024, hadfi_augmented_2022, chiang_enhancing_2024, reicherts_extending_2022}, creating opportunities to transform VAAs into more engaging discussion tools that inspire thoughtful effort. Exploratory work is being done to bring LLMs into the political domain, with tools ranging from politician avatars \cite{mancera_andrade_genai_2024} to retrieval-augmented generation (RAG) \cite{lewis_retrieval-augmented_2020} pipelines that seek to generate unbiased answers about political parties \cite{schiele_voting_2024}. We build on this initial work to explore the design space for LLM-powered VAA chatbots through an early system deployment.

Exploring the design space through empirical evidence is relevant for civic education, as there is also a gap in risk assessment for election-related use cases of LLMs. LLMs' tendency to produce non-factual information not only affects user satisfaction \cite{kim_understanding_2024} but also poses risks to election integrity. The risks have been underscored by numerous expert evaluations on election-related facts (e.g. candidates' names)  \cite{helming_generative_2023, romano_dataset_2024, angwin_seeking_2024, anthropic_testing_2024}, as well as automated evaluations revealing LLMs' supposed bias towards left-leaning policies based on the popular Political Compass test \cite{feng_pretraining_2023, hartmann_political_2023, motoki_more_2023} and existing European VAAs \cite{chalkidis_llama_2024,rettenberger_assessing_2024}. However, such system-based probing overlooks users' realistic input and reliance on the output.

\subsection{Trust in AI}
Trust in AI systems, such as LLM-based chatbots, hinges on the system’s trustworthiness and users’ attitudes toward it \cite{liao_designing_2022, laux_trustworthy_2024}. Trustworthiness in an LLM chatbot can be assessed based on its competence and value alignment with users \cite{manzini_should_2024}, ideally corresponding to appropriate levels of user reliance on the technology \cite{schemmer_appropriate_2023}. It is also crucial to avoid misplaced trust in an AI's capabilities, and several approaches have been proposed to limit overreliance on human-AI collaboration \cite{bucinca_trust_2021, lu_does_2024}. 

To communicate its trustworthiness, developers of an AI chatbot can use the MATCH mental model, consisting of the three components of underlying trustworthy \textit{attributes}, \textit{affordances} to communicate trustworthiness, and trust-related \textit{heuristics} for users \cite{liao_designing_2022}. The \textbf{attributes} of the MATCH model refer to the system’s core qualities, such as its intentions, fairness, and privacy standards. The \textbf{affordances} refer to a system's design features or interactive elements that enable users to perceive and act upon cues related to trustworthiness. The \textbf{heuristics} are mental shortcuts often invoked by effective cues of trustworthiness. Together, these components help designers deploy AI systems that are technically robust and ethically sound, which ultimately foster meaningful engagement. While the MATCH model has been applied to improve applications in healthcare \cite{jin_better_2024}, education \cite{ding_students_2023} and public services \cite{lopez_users_2024}, it is yet to be applied in a political domain. Through our work, we extend these principles to LLM-based VAA chatbots.

 \subsubsection*{In summary} Studies show that VAAs can improve political knowledge and voting intention, but significant knowledge prerequisites, uneven adoption, and narrow focus on rigid policy statements constrain VAAs' effectiveness. The limited scope of VAAs' intended outcome also invites reimagination of their functionality via interface design. LLM-based conversational interfaces offer the potential to enhance voter engagement through personalized and interactive explanations, yet there is a lack of empirical results from a realistic context to inform its design, and uncertainties remain about how users navigate LLMs' tendency to generate non-factual information and exhibit political bias. To address these gaps, our research investigates how voters can overcome the challenges of existing VAAs with an LLM-based chatbot interface, explores ways that it can unlock new opportunities, and identifies key challenges to be addressed. Lastly, we extend the MATCH model to the design of a trustworthy AI chatbot interface for civic education.

\section{Methods}

\subsection{Approach to Chatbot Design}

Since chatbot interfaces for voting preparation are only starting to appear, there has been no established design method for an LLM-based VAA chatbot. To lay the groundwork for future design, we observe how users engage with a VAA chatbot that generates voting advice in real time, extending the scope of existing work that uses a scripted chatbot as a separate explainer to a VAA \cite{kamoen_i_2022, van_zanten_voting_2024}. Similar to prior exploratory studies \cite{jordan_chatting_2023, reicherts_extending_2022} in the domain of conversational user interface (CUI), our work is more focused on understanding users' interaction styles and requirements than on refining specific interface features or fixing certain parameter choices.

We envisioned two broad interactive styles that prospective users would engage with a VAA chatbot---ChatGPT-like interaction involving open-ended questions and turn-by-turn opinion survey with policy statements evocative of existing VAAs---and specified the chatbot's behavior accordingly. Noting the continuously advancing state of the art in LLMs' architectures and efforts in enhancing truthfulness, our prototyping approach is intentionally model-agnostic, aiming to uncover users' perspectives on this interactive experience in civic education, which will remain relevant in the future.

\subsection{System design: LLM-Based Chatbot }

\subsubsection{Technical Implementation}
Our prototype chatbot runs on OpenAI’s GPT-4o model with custom prompts. We chose prompt engineering over fine-tuning or retrieval-augmented generation (RAG) to shape a specific type of dialogue without adding complexity that could affect participants’ perception of LLMs' truthfulness and bias. OpenAI’s model was selected for two reasons. First, ChatGPT is the most widely used LLM-based chatbot, and using a similar underlying model brings familiarity to participants and leverages affordances from prior usage to guide interaction. Second, thanks to its popularity, the GPT model family has been well-evaluated by safety researchers and its risk profiles are better understood than others’; OpenAI itself has also performed risk assessments with respect to political misinformation \cite{openai_gpt-4o_2024}.

\subsubsection{Interaction Design}
\label{sec:interaction_design}

\begin{figure*}[htbp]
    \centering
    \includegraphics[width=0.8\textwidth]{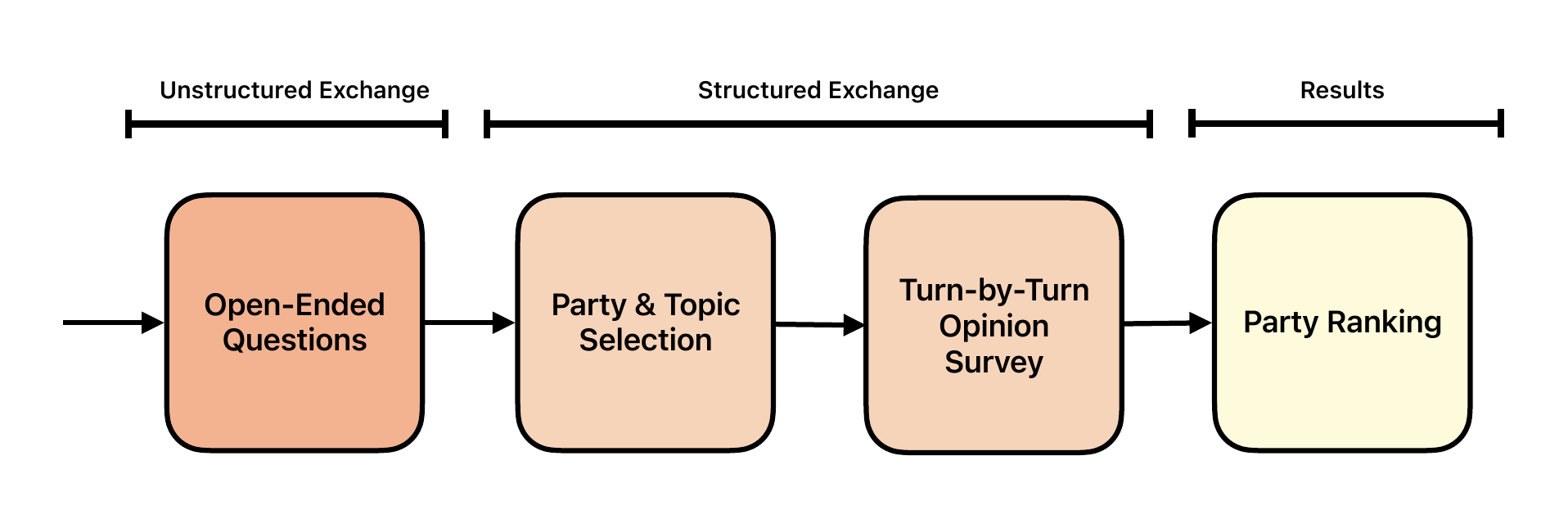}
    \caption{Conceptual walk-through of the interaction with the chatbot (as explained in section \ref{sec:interaction_design}).}
    \Description[the unstructured exchange, followed by the structured exchange]{On the left side of the flow chart, the icons and arrows indicate that the chatbot interaction starts with open-ended questions (unstructured exchange) followed by the structured exchange. The last arrow points to a box where a party ranking is provided to the user.}
    \label{fig:chatbot_interaction}
\end{figure*}

In order to evoke comparison with traditional VAAs while introducing greater interactivity, we incorporated both the turn-by-turn structure of traditional VAAs and a chatbot's invitation for an open-ended discussion, configured with our custom German prompts. We conceptually break down the chatbot interaction into \textbf{unstructured} and \textbf{structured} exchanges (as shown in Figure \ref{fig:chatbot_interaction}). In the unstructured exchange, users were invited to ask any election-related questions, prompted by the chatbot’s opening message (shown in Figure \ref{fig:screenshots}a). Once users had no further questions, the chatbot would guide them to the structured exchange, where they could select parties and topics of interest to be included in the subsequent turn-by-turn opinion survey. At each turn, users were asked to respond to a VAA-like statement with a clear stance (example in Figure \ref{fig:screenshots}b), and the chatbot would explain the alignment between users' stances and their selected parties' platforms. Similar to Wahl-O-Mat’s answer buttons, the chatbot requested stances on a three-point scale; unlike the rigid choices offered by buttons, the affordance of a text-input box allowed users to further elaborate and pose questions if they felt inclined. After 10 iterations by default, the chatbot would present a final ranked list of the selected parties. Our system prompts are included in Appendix \ref{sec:system_prompts}.

\subsection{Participant Recruitment and Demographics}

We recruited 331 Germany-based German citizens fluent in German via the research platform Prolific\footnote{\url{https://www.prolific.com/}}, known for its comparably higher response quality \cite{veselovsky_artificial_2023}. Of the 350 participants who signed up, 19 were screened out by attention checks. To approximate stratified sampling, we created concurrent Prolific tasks targeting six education-sex combinations, which limited over-representation without producing a truly balanced dataset. After the chatbot interaction and survey, we conducted follow-up interviews with 10 participants, who were recontacted and compensated through Prolific. Compensation exceeded the German minimum hourly wage for both the survey and interviews.

Participants had a median age of 30, and 48\% held at least a bachelor’s degree. The gender distribution was 37\% women and 2\% non-binary. Regarding usage of LLM-based chatbots, 49\% used one at least weekly, while 26\% had little to no prior experience. ChatGPT was the most popular chatbot (46\% weekly users), followed by Copilot (7\%). On attitude towards AI , 47\% believed AI was “good for society.” 88\% had used Wahl-O-Mat, Germany’s popular VAA---compared to an estimated adoption rate of 37\% among eligible voters\footnote{Based on 60.9 million eligible German voters, a 64.8\% turnout, and 14.8 million recorded uses of Wahl-O-Mat.
\\https://www.bundeswahlleiterin.de/info/presse/mitteilungen/europawahl-2024/06\_24\_wahlberechtigte.html 
\\https://www.bpb.de/themen/europawahlen/dossier-europawahlen/549597/rueckblick-die-europawahl-2024-in-deutschland-im-europaeischen-kontext}. These participant statistics are reproduced in Table \ref{tab:participant_stats}. 

Participants' political orientation skewed left, with a mean score of 3.8 (\(SD=2.1\)) on a 0 (extreme left) to 10 (extreme right) scale. They also showed stronger support for left-leaning parties than the general public (Table \ref{tab:participant_party_choices_table}).

\begin{table*}[htbp]
\centering
\begin{tabular}{|l|l|}
\hline
\textbf{QUESTION}          & \textbf{RESPONSE}                                          \\ \hline
\textbf{Total Responses}    & 331 (out of 350 who passed attention checks)                   \\ \hline
\textbf{Median Age}         & 30 years (\( IQR\ 24\text{-}37 \))                              \\ \hline
\textbf{Education Level}    & 48\% with at least a Bachelor’s Degree                             \\ \hline
\textbf{Gender}             & Women 37\%, Non-binary 2\%                                 \\ \hline
\textbf{Weekly LLM Chatbot Users} & 49\%  (used at least once a week)    \\ \hline
\textbf{Infrequent Chatbot Users}  & 26\% (never used or only a few times)               \\ \hline
\textbf{Most-Used Chatbots}  & ChatGPT 46\% (weekly usage), Copilot 7\%                   \\ \hline
\textbf{Attitude Towards AI} & 47\% “Good for society”, 13\% “Bad for society”          \\ \hline
\textbf{Wahl-O-Mat Usage}   & 88\% (compared to 37\% wider adoption rate)                \\ \hline
\end{tabular}
\caption{Characteristics of survey participants.}
\label{tab:participant_stats}
\end{table*}

\begin{table*}[htbp]
    \centering
    \begin{tabular}{lccc}
        \toprule
         Party & \makecell{Support Among \\Participants (\%)} & Vote Share (\%) & \makecell{Left-Right (0-10)\\ Orientation }  \\

        \midrule
        Grünen & 23.6 & 11.9 & 3.2 \\
        SPD & 11.2 & 13.9 & 3.6 \\
        Volt & 10.6 & 2.6 & - \\
        Die Linke & 7.3 & 2.7 & 1.4 \\
        AfD & 7.3 & 15.9 & 9.2 \\
        CDU/CSU & 6.9 & 30.0 & 5.9/7.2 \\
        FDP & 6.6 & 5.2 & 6.4 \\
        BSW & 6.0 & 6.2 & - \\
        Die PARTEI & 3.3 & 1.9 & - \\
        MERA25 & 3.0 & 0.3 & - \\
        Piraten & 2.4 & 0.5 & 2.1 \\
        Tierschutzpartei & 0.9 & 1.4 & 2.3 \\
        Freie Wähler & 0.9 & 2.7 & - \\
        None Selected & 10.0 & &  \\
        \bottomrule
    \end{tabular}
    \caption{Ranked list of participants' most favored parties. The \textit{Vote Share} column reflects the eventual German outcome of the 2024 European Parliament election\protect\footnote{ https://www.bundeswahlleiterin.de/en/europawahlen/2024/ergebnisse/bund-99.html}. The parties' left-right political orientations are taken from the Chapel Hill Expert Survey \cite{jolly_chapel_2022}.}
    \label{tab:participant_party_choices_table}
\end{table*}
\footnotetext{\url{ https://www.bundeswahlleiterin.de/en/europawahlen/2024/ergebnisse/bund-99.html}}

\subsection{Study Procedure and Tasks}
\label{sec:study_procedure}

\begin{figure*}[htbp]
    \centering
    \includegraphics[width=0.8\textwidth]{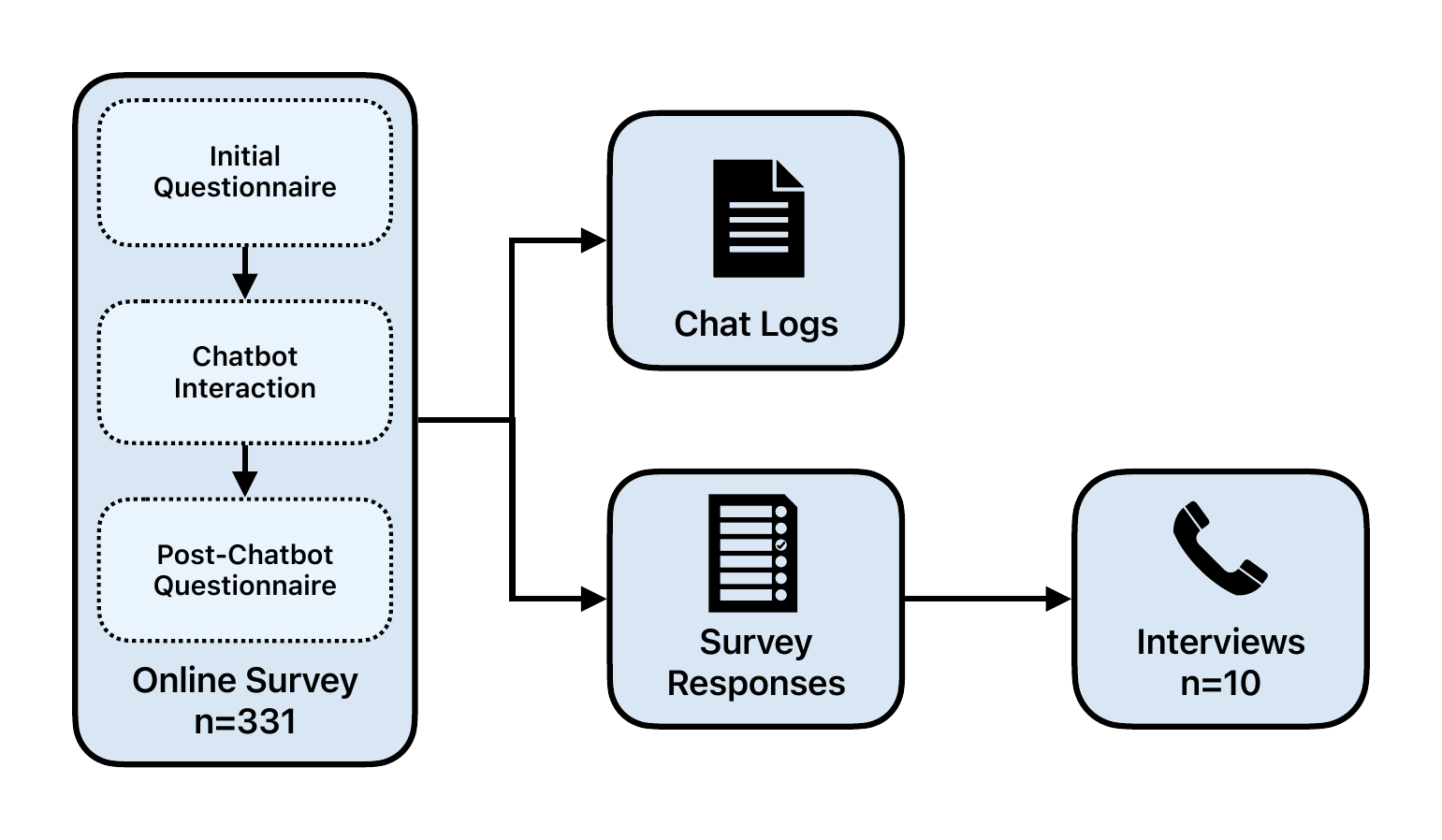}
    \caption{Overall flow of the study (as explained in Section \ref{sec:study_procedure}).}

    \Description[Survey with chatbot, followed by interviews]{On the left side of the flow chart, the icons and arrows indicate 331 participants taken from an initial questionnaire, to a chatbot and another questionnaire. Arrows point to the center from the left, at "survey responses" and "chat logs". Then the arrows point to the right, an icon for interviews involving 10 participants.}
    \label{fig:study_procedure}
\end{figure*}

The two-phase study procedure consists of a survey and follow-up interviews to collect quantitative and qualitative data. It is mapped out in Figure \ref{fig:study_procedure} and expanded in the following subsections.

\subsubsection{Chatbot Interaction and Survey}
Participants first provided informed consent and agreed to the prohibition of AI use at the beginning of the Qualtrics\footnote{\url{https://www.qualtrics.com/strategy/research/survey-software/}} survey. To emulate the prevailing modality of chatbot interaction and reduce friction to typing, they were required to access the chatbot via a desktop web browser, the most common way users interact with ChatGPT\footnote{Over 60\% of web traffic to ChatGPT’s domain came from desktop devices in May 2024, according to \url{https://www.similarweb.com}.}.  

\noindent The survey consisted of three tasks:  

\begin{enumerate}
    \item \textbf{Initial Questionnaire}: Participants answered demographic questions (e.g., age, gender, education) and provided information on their prior experience with LLM chatbots, opinions on AI’s societal role, and political attitudes.  
    \item \textbf{Chatbot Interaction}: Participants engaged with the chatbot via an embedded web app following the interaction, as outlined in Section \ref{sec:interaction_design}. Chat logs were recorded for analysis.  
    \item \textbf{Post-Chatbot Questionnaire}: Participants were asked to reassess their political attitudes and provided feedback on chatbot usability, voting intention, and output quality. Survey items were adapted from the Chatbot Usability Questionnaire (CUQ) \cite{holmes_usability_2019} and the NASA Task Load Index (NASA-TLX) \cite{hart_nasa_1986}. Most questions in the survey were on a Likert scale, with some followed by an optional text field to solicit explanation and encourage thoughtful selection. The questions and response scales are shown in Appendix \ref{sec:appendix_survey}.
\end{enumerate}

\noindent Before exiting, participants received a disclaimer on the chatbot’s limitations and potential AI-generated misinformation. The English translation of the original German survey is provided in Appendix \ref{sec:appendix_survey}.  

Participants had a median chatbot session time of 606 seconds (\(M=736\), \(SD=486\)), sending a median of 15 messages (\(M=15.9\), \(SD = 5.3\)). 39\% reached the end of the structured exchange by responding to all 10 policy statements.

\subsubsection{Follow-Up Interviews}

To complement our survey findings, we conducted 10 follow-up interviews to gain deeper insights into user experiences. With the goal of capturing diverse perspectives, we selected 10 participants with varying attitudes towards AI, political dispositions, and opinions of the chatbot experience (summary in Table \ref{tab:interviewee_stats})

We conducted 30-minute, audio-only interviews in German via Microsoft Teams, starting the recording upon verbal reaffirmation of consent. Transcripts were created with AssemblyAI\footnote{\url{https://www.assemblyai.com/}} and translated into English using DeepL\footnote{\url{https://www.deepl.com/}}.

\begin{table*}[htbp]
    \centering
    \resizebox{\textwidth}{!}{%
    \begin{tabular}{l|cccc|ccc|cccc}
    \hline
    &\multicolumn{4}{c|}{\textbf{Demographics}} & \multicolumn{3}{c|}{\textbf{Initial Questionnaire}} & \multicolumn{4}{c}{\textbf{Post-Chatbot Questionnaire}} \\ \hline
    \makecell{\textbf{Interviewee} \\ \textbf{Code}} & \textbf{Age} & \textbf{Gender} & \makecell{\textbf{Education}} & \makecell{\textbf{Attitude} \\ \textbf{towards}\\ \textbf{AI}} & \makecell{\textbf{Political} \\ \textbf{Interest}  \\ \textbf{(Lo-Hi: 1-4)}} & \makecell{\textbf{Political} \\ \textbf{Self-Efficacy} \\ \textbf{(Lo-Hi: 1-4)}} & \makecell{\textbf{Political}  \\\textbf{Orientation} \\ \textbf{(L-R: 0-10)}}& \makecell{\textbf{Reuse} \\ \textbf{Intention} \\ \textbf{(Lo-Hi: 1-5)}} & \makecell{\textbf{Perceived}  \\\textbf{Chatbot} \\ \textbf{Accuracy} \\ \textbf{(Lo-Hi: 1-7)}}& \makecell{\textbf{Perceived} \\ \textbf{Knowledge} \\ \textbf{Gain} \\ \textbf{(Lo-Hi: 1-7)}} & \makecell{\textbf{Change in} \\ \textbf{Political Interest}} \\ \hline \hline
    
    P1 & 34 & Male & Bachelor & Neutral & 3 & 2 & 3 & 5 & 5 & 6 & Increased \\ \hline
    P2 & 34 & Male & Master & Positive & 4 & 4 & 2 & 2 & 6 & 1 & Unchanged \\ \hline
    P3 & 45 & Non-Binary & High School & Negative & 3 & - & 0 & 5 & 4 & 4 & Unchanged \\ \hline
    P4 & 21 & Male & High School & Positive & 3 & 2 & 6 & 5 & 7 & 6 & Decreased \\ \hline
    P5 & 53 & Male & Bachelor & Positive & 4 & 3 & 8 & 4 & 7 & 5 & Unchanged \\ \hline
    P6 & 27 & Female & High School & Positive & 3 & 3 & 1 & 4 & 3 & 3 & Unchanged \\ \hline
    P7 & 21 & Male & High School & Positive & 3 & 4 & 4 & 5 & - & 6 & Increased \\ \hline
    P8 & 28 & Female & Master & Negative & 2 & 2 & 4 & 5 & 6 & 6 & Unchanged \\ \hline
    P9 & 24 & Female & High School & Negative & 2 & 1 & 1 & 4 & 6 & 6 & Unchanged \\ \hline
    P10 & 31 & Male & Master & Neutral & 4 & 3 & 8 & 2 & 5 & 1 & Unchanged \\ \hline
    \end{tabular}%
    }
    \caption{Interviewees and their key characteristics.}
    \label{tab:interviewee_stats}
\end{table*}
The interviews were semi-structured and invited participants to talk about their civic engagement, political participation, media diet, trust in information sources, voting preparation strategy, informedness, trust in our chatbot, and opinions about AI’s role in educating voters.

\subsection{Details on Survey items}

\subsubsection{Usability and Taskload} To benchmark the overall user experience of our chatbot, we included questions from the Chatbot Usability Questionnaire (CUQ) \cite{holmes_usability_2019}, which is designed to be comparable to the established usability metric System Usability Scale (SUS) \cite{brooke_sus_1996}. Similar to SUS, the responses were converted to a score out of 100. To evaluate participants' cognitive workload, we asked the questions of Mental Demand, Performance and Frustration from the NASA Taskload Index (NASA-TLX) on a 7-point scale.

\subsubsection{Political Attitude}
We asked participants to report their political orientation on a 0-10 left-right scale (with 0 being extreme left and 10 extreme right) common in expert surveys \cite{jolly_chapel_2022} and solicited their levels of interest in politics and self-efficacy in political participation (1-4 scale). To capture any explicit change in intention as a result of chatbot interaction, we also asked participants both before and after the chatbot session their inclination to vote for each of the seven most popular parties and any other party they specified (0-10 scale).

\subsubsection{Perceived Effects of the Chatbot} After chatbot use, we asked participants to self-report political knowledge gain and voting intention on a 7-point Likert scale, using questions similar to those of Kamoen and Liebrecht \cite{kamoen_i_2022}.

\subsubsection{Perceived Truthfulness and Partisan Biases} Participants were asked to report their perceptions of the truthfulness and partisan biases in the chatbot’s responses. One question focused on the accuracy of the information regarding various political perspectives, using a 7-point scale to gauge participant responses. Another question examined perceptions of fairness in how the chatbot presented and compared political parties. Participants were asked to indicate the extent to which the chatbot’s responses favored certain parties.

\subsection{Data Analysis}

\subsubsection{Quantitative}

To assess how demographic and user characteristics influenced responses to Likert-scale survey questions, we fitted ordinal regression models using the Python \textit{statsmodels} library. We applied the Benjamini-Hochberg correction \cite{benjamini_controlling_1995} to control the false discovery rate and set statistical significance at $\alpha = 0.05$. This analysis examined the impact of factors such as age, education, and political interest on user perceptions of the chatbot.

To analyze user queries from the unstructured exchange, we categorized questions into five broad types informed by prior evaluations \cite{romano_dataset_2024, anthropic_testing_2024}. We used GPT-4o for zero-shot classification \cite{ziems_can_2024}, excluding non-question inputs labeled as “no question.” Definitions and classification prompts are provided in Appendix \ref{sec:appendix_zeroshot}.

\subsubsection{Qualitative}
We analyzed interview transcripts and written survey responses using Braun and Clarke’s thematic analysis approach \cite{braun_using_2006}. Two researchers independently reviewed and coded the data in multiple passes, ensuring equal attention to all excerpts. They then compared notes, resolving discrepancies through discussion. The research team synthesized the final themes in relation to our research questions.

\subsection{Ethics and Positionality}

This study was approved by the ethical review board of Saarland University’s Computer Science Department (ID: 24-05-7). No personally identifiable information (PII) was collected, and algorithmic PII scrubbing was performed before analysis. We employed established cloud-based services to process the data; Prolific, Qualtrics, AssemblyAI and DeepL all reported processing their EU customers’ data within the bloc. The study took place the week before the European Parliament election in Germany to engage participants with genuine electoral interest. To prevent undue influence, the post-chatbot survey cautioned participants against relying on the chatbot responses as voting advice and encouraged them to visit official election resources. A follow-up message before the election day reiterated the risks from AI-generated misinformation and encouraged independent research.

Our team comprises researchers in computer and political sciences, based in a Western European country, with experience conducting internet-mediated research across Europe. Team members are native speakers of multiple languages, including German. Our study is driven by a commitment to understanding the role of LLMs in civic education for German voters, shaping both our research design and analysis.

\section{Findings}
We group the findings into improvements to VAAs enabled by LLM-based chatbots (RQ1), opportunities beyond the scope of traditional VAAs (RQ2), and obstacles to be overcome in future design processes (RQ3). We use the term "participants" to refer to chatbot users who completed the study, and use "interviewees" (and the codes P1-P10 shown in Table \ref{tab:interviewee_stats}) to attribute insights to interviewees.

\subsection{Clear and Concise: Improving VAA Accessibility and Effectiveness with LLM-Based Chatbots (RQ1)}
\label{sec:findings_rq1}

Participants appreciated how the conversational abilities of LLMs made the VAA interaction more accessible by lowering the hurdles of complex terminology and prerequisite knowledge, reporting high satisfaction with our prototype chatbot.

\subsubsection{Transforming Complex Political Information into Concise and Accessible Answers}

Interviewees described the experience of reading the results from Wahl-O-Mat---the popular German VAA---as overwhelming, with one equating it to walking into "\textit{a wall of text}" (P6). As a result, another noted that they "\textit{only ever looked at these percentages}" (P7), referring to the numeric alignment score that Wahl-O-Mat produces for each party of interest, and would rarely read more than a few sentences. In contrast, interviewees valued the chatbot's ability to produce concise and personalized answers, which helped provide an overview of the issues and parties of interest. "\textit{It was really like talking or writing to a person who knows about it and just gives you some kind of advice in that direction}" (P4), said a young interviewee of the relevance and agreeableness of the answers. 

In agreement with prior work \cite{kamoen_i_2017}, participants also reported feeling uncertain about the meaning of some statements when using Wahl-O-Mat. On the tone and complexity of the language, participants overwhelmingly preferred the chatbot, with an interviewee elaborating that “\textit{the wording was friendly}” and "\textit{the answers were of a pleasant length}” (P5), which made the information more accessible to consume.

More than half of the interviewees viewed the chatbot as an easy way of seeking information about political parties. With the chatbot, "\textit{people can get the information they want very quickly and very informatively}" (P7).

\subsubsection{Offering On-demand Clarification}

\label{sec:versatility} 
Participants frequently highlighted the VAA chatbot's flexibility in handling various requests. An interviewee found it helpful that they could ask follow-up questions---in the form of "\textit{what do you mean by XY?}" (P5)---at any point and have new concepts clarified as they were brought up. Others appreciated the chatbot's ability to produce concise two-way comparisons between parties of interest. For an interviewee (P7) who had a rough idea of what each party stood for beforehand, the chatbot's ability to explain the subtle difference between two parties' positions in one sentence "\textit{sparked interest}" (P7) in continuing to learn from the chatbot. The chatbot's ability to answer any question on demand, which traditional VAAs lack, contributed to the feeling of being informed.

\subsubsection{Easy to Navigate}

The chatbot demonstrated high usability, with a median CUQ score of 84 (\( M=80.9, SD=14.0 \)), placing it in the 90-95th percentile against an SUS benchmark \cite{lewis_item_2018}. They also found the chatbot effective in performing the task of voting preparation with medium effort and low friction. On NASA-TLX, participants reported median ratings of 3 for \textit{Mental Demand} (\( IQR\ 2\text{-}5 \)), 6 for \textit{Performance} (\( IQR\ 5\text{-}6 \)), and 1 for \textit{Frustration} (\( IQR\ 1\text{-}2 \)) on a 7-point scale. Most participants expressed a willingness to use the chatbot again (\( Md=4, IQR\ 3\text{-}5 \); 5-point scale) and to recommend it to a friend (\( Md=4, IQR\ 2\text{-}5 \)).

\subsubsection{Informative and Well-Received}

Participants generally reported an increase in political knowledge through the chatbot interaction, largely agreeing that they had "gained more understanding of the political landscape", with a median rating of 5 on a 7-point scale (\( IQR\ 4\text{-}6 \)). They also felt more motivated to vote (\( Md=5, IQR\ 4\text{-}6 \)).

Using different combinations of user characteristics, we fitted four ordinal regression models for the two response variables of interest: perceived political knowledge gain (Table \ref{tab:regression_tabel_knowledge}), voting intention, reuse intention and bias perception (Appendix \ref{sec:appendix_regression}) after chatbot interaction. For each regression, Model 1, the best-fit model by AIC score, uses education, political interest, political orientation, and attitude towards AI as predictors. Model 2 differs from Model 1 by using political self-efficacy instead of political interest (Spearman's $\sigma=0.573$). Model 3 replaces the AI attitude in Model 1 with experience using existing LLM-based chatbots (Spearman's $\sigma=0.324$). Model 4 includes the variable age group, which is commonly associated with VAA and technology use. 

The regression coefficients show that perceived effects were mediated by certain demographic and behavioral traits. Positive attitude towards AI and frequent use of LLM chatbot both had a strong positive correlation with knowledge gain. Compared to those who rarely or never used an LLM chatbot, weekly and daily users were more than twice as likely to report a higher level of political knowledge gain (\(Odds\: Ratio=2.42, p=0.01\) for daily users) (Table \ref{tab:regression_tabel_knowledge}). Participants with lower educational attainment were better informed by the chatbot than the reference group with an advanced degree (master or above), with those without a degree being nearly twice as likely to report higher knowledge gain (\(OR=1.97, p=0.02\)). Regarding levels of political engagement, participants with moderately high (3) political interest (\(OR=2.01, p=0.04\); highest level (4) as reference) and moderately low (2) political self-efficacy (\(OR=2.49, p=0.04\); highest level (4) as reference) were most likely to report higher knowledge gain. Apart from attitude towards AI, we saw no statistically significant effect of user characteristics on voting intention (Table \ref{tab:regression_tabel_vote} in Appendix \ref{sec:appendix_regression}).

Political orientation is a clear mediating factor in participants' opinions of the chatbot, which may indirectly affect the tool's educational outcome. The further right a participant identified on the political spectrum (continuous variable), the more likely they were to perceive bias in the chatbot ($OR=1.32, p = 0.01$; Table \ref{tab:regression_tabel_bias} in Appendix \ref{sec:appendix_regression}). Similarly, the political orientation may have a marginal effect on reuse intention ($OR=0.89, p = 0.07$; Table \ref{tab:regression_tabel_reuse} in Appendix \ref{sec:appendix_regression}).

\begin{table*}[htbp]
\resizebox{\textwidth}{!}{
    \begin{tabular}{l|ll|ll|ll|ll}
    \toprule
    Outcome Variable: Political Knowledge Gain (7-point Likert scale) & \multicolumn{2}{c|}{\textbf{Model 1}} & \multicolumn{2}{c|}{Model 2} & \multicolumn{2}{c|}{Model 3} & \multicolumn{2}{c}{Model 4} \\
     & \textbf{Odds Ratio} & \textbf{Standard Error} & Odds Ratio & Standard Error & Odds Ratio & Standard Error & Odds Ratio & Standard Error \\
    \midrule
    Age Group (ref: 50+) 18-24 &  &  &  &  &  &  & 1.713   & 1.723 \\
    Age Group (ref: 50+) 25-34 &  &  &  &  &  &  & 1.702   & 1.679 \\
    Age Group (ref: 50+) 35-49 &  &  &  &  &  &  & 1.675   & 1.709 \\
    Education Level (ref: Master or Above) High School or Below & \textbf{1.969*}  &\textbf{1.298} & 1.921*  & 1.300 & 2.155*  & 1.297 & 1.973   & 1.335 \\
    Education Level (ref: Master or Above) Bachelor & \textbf{1.210}   & \textbf{1.343} & 1.257   & 1.346 & 1.422   & 1.335 & 1.246   & 1.347 \\
    Political Interest (ref: 4) 1 & \textbf{0.673}   & \textbf{1.742} &  &  & 0.823   & 1.737 & 0.677   & 1.758 \\
    Political Interest (ref: 4) 2 & \textbf{1.710}   & \textbf{1.369} &  &  & 1.840   & 1.377 & 1.640   & 1.379 \\
    Political Interest (ref: 4) 3 & \textbf{2.014*}  & \textbf{1.343} &  &  & 2.172*  & 1.349 & 1.956   & 1.349 \\
    Political Self-Efficacy (ref: 4) 1 &  &  & 1.519   & 1.730 &  &  &  &  \\
    Political Self-Efficacy (ref: 4) 2 &  &  & 2.489*  & 1.468 &  &  &  &  \\
    Political Self-Efficacy (ref: 4) 3 &  &  & 1.714   & 1.476 &  &  &  &  \\
    Political Orientation (Left-Right: 0-10 continuous) & \textbf{0.924}   & \textbf{1.052} & 0.936   & 1.052 & 0.945   & 1.052 & 0.932   & 1.053 \\
    Attitude towards AI (ref: Negative) Neutral &\textbf{1.236}   & \textbf {1.388} & 1.098   & 1.388 &  &  & 1.235   & 1.391 \\
    Attitude towards AI (ref: Negative) Positive & \textbf{3.119**} &\textbf{1.388} & 2.683*  & 1.388 &  &  & 3.085** & 1.394 \\
    Chatbot Usage (ref: Rarely) Monthly &  &  &  &  & 1.587   & 1.323 &  &  \\
    Chatbot Usage (ref: Rarely) Weekly &  &  &  &  & 2.230*  & 1.318 &  &  \\
    Chatbot Usage (ref: Rarely) Daily &  &  &  &  & 2.424*  & 1.383 &  &  \\
    AIC & \textbf{1077.310} &  & 1078.974 &  & 1089.953 &  & 1082.198 &  \\
    \bottomrule
    \multicolumn{9}{c}{$*p<0.05, **p<0.01$ \hspace{0.5cm} The best-fit model (lowest AIC score) is in bold.} \\
    \end{tabular}}
    \caption{Estimates (odds ratios) from ordinal regression models predicting perceived political knowledge gain by user characteristics.}
    \label{tab:regression_tabel_knowledge}

\end{table*}

\subsection{Curiosity, Reflection and Rationalization: New Opportunities of LLM-Based VAAs (RQ2)}

Participants valued the curiosity-driven exploration and reflection on their own opinions facilitated by the conversational interaction, which helped rationalize electoral choices.

\subsubsection{Turn in Any Direction: Chatbot’s Flexibility in Answering Various Types of Questions}

Interviewees found it helpful that the chatbot could offer more detailed information when they were curious about a topic it introduced. One interviewee recalled that the chatbot would ask "\textit{Do you want to discuss the topic in more depth or move on to the next question?}" (P8) at the end of each round, feeling reassured that their desire for further exploration could always be fulfilled.

The chatbot's ability to handle open-ended questions was particularly useful for interviewees who struggled to find information elsewhere. One interviewee (P2), who had a high interest in politics but found it challenging to phrase effective Google search queries, found the chatbot more useful for exploring topics. A breakdown of questions participants asked during the unstructured exchange is shown in Table \ref{tab:unstructured_questions_cat_table}.

Another (P9), who had been overwhelmed by the vast amount of anxiety-inducing information on public media, felt empowered by the chatbot's ability to steer the conversation to discuss precisely what they were curious about.

\begin{quote}
"\textit{I think it's quite good that you can turn in any direction and perhaps ask more questions, or take a closer look at a party or have an overall view and direct comparison.}" (P9)
\end{quote}

\begin{table*}[htbp]
    \centering
        \begin{tabular}{lrrl}
        \toprule
         Type & Share (\%) & Count & Example\\
        \midrule
        Party-specific & 31.0 & 172 & "What does the CDU stand for?"\\
        Irrelevant & 24.9 & 138 & "Hello, how are you today?"\\
        Topic-specific & 23.4 & 130 & "Which parties support e-mobility?"\\
        General EU & 13.5 & 75 & "How many members does the European Parliament have?"\\
        Administrative & 3.4 & 19 & "How can I vote by mail?"\\
        Miscellaneous & 3.1 & 17 & "How should I decide whom to vote for?" \\
        Regional & 0.7 & 4 & "Where can I vote in Berlin-Mitte?"\\
        \midrule
        Total & & 555 & \\
        \bottomrule
        \end{tabular}
    \caption{Types of user questions in the unstructured exchange.}
    \label{tab:unstructured_questions_cat_table}
\end{table*}

\subsubsection{Encouraging Reflection and Rationalization}
\label{sec:findings_reflection}

Interviewees appreciated being prompted to reflect on their understanding and opinions. One (P9) described feeling encouraged to think in "\textit{a slightly different and deeper way}" about specific topics and consider "\textit{what certain problems might be and how they could be tackled}". Another (P4), recalling the experience of reading two sides of the argument, reported having actively asked themselves "\textit{what's more important}", which was a more rewarding cognitive activity than passive reading.

Interviewees felt the in-depth discussion introduced them to topics they didn’t understand and helped rationalize their electoral choices. One (P8) recognized that unlike a traditional VAA that only invites users to click a button after reading about an issue, the chatbot encouraged a dialogue at every turn, prompting them to think and reflect.

\begin{quote}
"\textit{It was quite good because you simply had to reason more...you couldn't just say, yes...you had to think: okay, why am I actually for or against it?}" (P8)
\end{quote}

The chatbot gave interviewees an overview of party platforms and reinforced their political knowledge. Interviewees described the chatbot as useful for consolidating their understanding of political parties. One (P3) noted that "\textit{it hasn't changed my opinion, but it has helped me categorize parties.}" Another (P8), who felt highly informed by the chatbot, used the exchange to reflect on topics they had limited knowledge about; the information provided rational grounding to their existing opinions, enhancing their sense of being informed.

Regarding the outcome of reflection prompted by the chatbot, however, mixed perceptions make it unclear whether such discussions can encourage users to consider new perspectives beyond reinforcing prior opinions. An interviewee reported an absence of opposing viewpoints in their discussion, remarking that "\textit{it didn't try to push me in any direction}" (P3). Other participants have both expressed concern over the chatbot echoing their exact opinions and praised its counterarguments that made them more open-minded.

\subsection{Unreliable, Opaque and Directionless: Obstacles to Trusted Utilization (RQ3)}

Participants expressed awareness of and concern about LLMs' known limitations in truthfulness and neutrality, emphasizing the importance of trustworthy mechanisms and cautioning against undue reliance. Nonetheless, some base their trust in the prototype chatbot on capabilities, drawing attention to the risk of misplaced trust. 

\subsubsection{Awareness of LLMs’ limitations}
\label{sec:findings_awareness}

Many interviewees expressed a strong awareness of potential risks associated with the use of LLMs as an information source, sharing their understanding of the technology's limitations from prior experience with tools such as ChatGPT.

\begin{quote}
"\textit{With ChatGPT, it was often the case that it gave me false information...it explained something mathematical to me that was only half right in the end. That's why I don't think I would generally trust chatbots 100\%.}" (P8)
\end{quote}

They similarly cautioned against overreliance on imperfect technology in making high-stakes decisions. As P5 summarizes \textit{“I would never blindly rely on it, especially when it comes to things like elections.”} (P5). Another (P7) revealed their general distrust of information from digital sources, explaining how they always consult multiple sources to verify new information. 

\begin{quote}
"\textit{I would say that a chatbot is very risky if it's the only knowledge medium.}" (P7)
\end{quote}

There was a consensus that a chatbot should be positioned as a tool to help voters begin---as opposed to conclude--- voting preparation. Many also suggest that users of voting preparation tools are responsible for double-checking everything they learn. As one put it, "\textit{you have to be aware of the risk, and ultimately you have to do your own research}" (P5).

\subsubsection{Desire for Traceability and Transparency}
\label{sec:findings_traceability-transparency}

All but one of the interviewees expressly stated their desire to see references in the chatbot’s responses, highlighting the importance of traceability to primary sources as a key to building trust. The ability to locate primary sources helps users double-check information, satisfying their curiosity.

\begin{quote}
"\textit{I would actually say that people who are really interested in it can look it up directly...if you don't trust the bot, you can read it again.}" (P4)
\end{quote}

Although interviewees found no obvious errors from the chatbot, some were reluctant to trust it because of the opacity in how the answers were generated.

\begin{quote}
"\textit{I think a chatbot like this is fed with training data, and I don't know what it has been given as training data...in other words, what it has been presented with.}" (P5)
\end{quote}

Several interviewees described neutrality as another prerequisite of trust for political information. An interviewee wanted information about the vested interests of the people behind the chatbot and would base their trust "\textit{on whether I think it's independent information or whether they have some kind of bias in it}" (P2). To assess and communicate the degree of neutrality of a chatbot, one (P7) suggested that it undergo independent audits by third parties. 

\begin{quote}
"\textit{It would be important for something like that to be checked somehow by some company and then certified in such a way that it is somehow neutral.}" (P7)
\end{quote}

Beyond behaviors that demonstrate neutrality and reliability, participants expressed eagerness to see proof of such qualities through transparency measures.

\subsubsection{Trust based on Perceived Reliability of Chatbot}
\label{sec:findings_trust-reliability}

While most interviewees demonstrated some a priori awareness of an LLM-based chatbot's limitations, few consequently questioned the veracity of the chatbot's claims.

The majority of survey participants (56.1\%) found the chatbot's presentation of various perspectives largely accurate, giving it a score of 6 or above on a 7-point scale. Most respondents also viewed the chatbot’s presentation and comparison of political parties as balanced. However, a minority perceived bias: 9.9\% felt the chatbot portrayed certain parties somewhat more positively, and 1.6\% believed it portrayed some parties much more favorably.

Interviewees described the chatbot’s outputs as plausible and neutral, and no one recalled specific examples of falsehood or bias. When asked about the trustworthiness of the chatbot, most---including those with a negative attitude towards AI---indicated a certain level of trust. 

Interviewees trusted the chatbot primarily due to its ability to communicate appropriately, offer reliable recommendations, and maintain a professional tone. They valued its capacity for facilitating meaningful exchanges of ideas and presenting coherent, sensible arguments. Additionally, its "\textit{scientific}" tone of writing, as noted by Interviewee P9, further enhanced its credibility. The chatbot also provided detailed responses and party recommendations that aligned with interviewees’ previous voting preparation experiences, solidifying it as a trustworthy tool in their view.

One interviewee (P3) with no prior usage of LLMs trusted the chatbot based on its appearance of competence---despite reporting a negative attitude towards AI in the survey. They had accepted the chatbot's recommendations at face value until being reminded of the potential inaccuracies. They then noticed warning signs that would have made them question the chatbot's trustworthiness, such as the echoing of their own opinions and the lack of disagreement throughout the interaction.

\subsubsection{Desire for User Guidance and Flexible Engagement}
\label{sec:findings_guidance-flexibility}

With regard to potential challenges in navigating such an interface, an interviewee (P10) commented that the chatbot's invitation to ask open-ended questions in the unstructured exchange could make users feel lost, especially if they have little idea what questions to ask to fill the knowledge gap.

\begin{quote}
"\textit{I imagine that most people are relatively uninformed...you can ask the chatbot questions, but you have to know what questions you want to know so that you know what you want to know.}" (P10)
\end{quote}

While interviewees appreciated instances when the chatbot asked whether they wished to "\textit{discuss the topic in more depth}" (P8), they would have valued more flexibility in deciding the depth of engagement with the chatbot. An interviewee with high political interest and self-efficacy(P2) found the repeated invitations to engage in a discussion repetitive, given the research they had already done before the study, and felt "\textit{rather overwhelmed}" (P2). Some survey responses suggested adding more GUI elements to streamline the responses, such as a button to indicate high interest in a particular topic.

\section{Discussion}

Our chatbot implementation yielded a very high CUQ score, demonstrating its user-friendliness. We also observed participants’ willingness to use it again, even among interviewees skeptical of AI's societal roles. Additionally, the chatbot yielded significant perceived knowledge gain among non-experts. These users, who tend to have lower educational levels, less confidence in political decision-making, and no particular interest in politics, are often underserved by traditional VAAs \cite{van_de_pol_beyond_2014, garzia_voting_2019, schultze_effects_2014}. This suggests that the chatbot can potentially engage a broader audience and make political information more accessible.

In addition to making an existing tool more accessible, the chatbot interface took the already familiar form of VAAs and transformed it into an engaging new tool eliciting greater cognitive effort. For users who were eager to learn, the experience of using a VAA chatbot was less like that of policy shoppers \cite{fossen_whats_2014}---following a well-trodden path from the entrance to the checkout lines---but one characterized by curiosity-driven exploration and reflection at their own pace. Breaking with the trend of AI use cases that reduce users’ critical thinking \cite{lee_impact_2025}, our deployment showcased an approach to redesign an existing educational tool with LLM to increase cognitive effort while improving user experience.

In contrast to concerns among civic organizations that LLMs may mislead voters, we uncovered participants' shrewdness with the technology and optimism in LLMs' role in civic education despite a shared awareness of risks. This can be explained by participants' inclination to regard VAAs---and a VAA chatbot---as a tool for preliminary research and the conviction that voters are responsible for doing their diligent research. Thus, participants see the potential for a chatbot to help them become better informed for elections and are willing to embrace an imperfect tool, provided that it keeps them engaged and meets their requirements for trust.

While some have feared the prospect of mass persuasion campaign enabled by LLMs \cite{goldstein_how_2024, bai_artificial_2023, durmus_measuring_2024, hackenburg_evaluating_2024}, we noted danger at the other extreme due to LLMs' tendency towards "sycophancy" \cite{sharma_towards_2023}, as participants reported having their opinions reinforced more often than being challenged. Sycophancy raises the risk that well-intentioned chatbots designed to facilitate critical thinking could---through their ability to pick up and emulate ideological language \cite{bleick_german_2024}---inadvertently lock the users in an echo chamber. Designers of future LLM-mediated discussion tools may benefit from an exploration of the trade-off between the competing ends of value alignment and critical thinking. On the one hand, blatant promotion of certain partisan views is unpalatable to the general public, and a discussion chatbot needs to take a pluralistic view to respect and personalize to the values of the users. On the other hand, excessive personalization can deprive users of the opportunity to recognize the weaknesses of their arguments, and a chatbot must constructively challenge users on contentious issues and offer diverse perspectives of reality. 

Germany, being the most populous EU state with a proportional representation system, is an archetype of a multi-party democracy, for which VAAs are most relevant. We hope our findings will pave the way for studies in other national contexts. We discuss design implications and establish future research directions in the following subsections.

\subsection{Recommendations for Fostering Effective Interaction with LLM-Based Chatbot for Civic Education}

\subsubsection{Probing Question for Active Reflection.}
Our interview provided initial evidence of a VAA chatbot's utility in facilitating reflection and deliberation (Section \ref{sec:findings_reflection}), yet despite the initial chatbot message inviting users to ask questions throughout the conversation, not all participants realized the chatbot's capability to respond to user messages containing more than just the response particles. Hence, users should be reminded of the possibility of in-depth discussions with probing questions and additional context. Simple "why" questions should be asked in response to contrarian views to encourage reflection, and hints of uncertainties should be responded to with offers of explanations. 

To help broaden users' perspectives, a chatbot can present multiple viewpoints through counterarguments or mini-debates to encourage articulation of their understanding and assessment of their assumptions. As we observed a challenge in generating consistently conservative arguments with LLMs, a self-fine-tuning method \cite{taubenfeld_systematic_2024} may be helpful in anchoring a model's political leaning, and new crowd-sourced datasets for political arguments would be helpful in national contexts underrepresented in LLMs pre-training data.

\subsubsection{Customization and Guidance.}
The chatbot format has successfully made VAA accessible (Section \ref{sec:findings_rq1}) to our participants, but the interface could be made more customizable to improve the effectiveness for users with varying needs. Different participants have pointed to the same chatbot features as both benefits and sources of irritation (Section \ref{sec:findings_guidance-flexibility}). Some described the in-depth discussions with the chatbot as a highlight of the experience, yet others who were disinterested in or already informed about a topic felt overwhelmed by the presentation of information. While some appreciated the ability to ask open questions, others felt lost when invited to ask questions without context. To engage users with varying levels of knowledge and interest, a customizable chatbot should allow users to specify engagement style, such as verbosity, and adapt dynamically to their revealed preferences. The chatbot should also offer detailed guidance about ways of interacting and customize the depth of discussions. An expandable information box or a tooltip may be helpful.

\subsubsection{Multimodal Interaction.}
Exploration of GUI elements and input modalities may improve user engagement. Some participants preferred additional interface elements like buttons (Section \ref{sec:findings_guidance-flexibility}) to streamline the interaction. Some prospective users may also prefer an audio-based interaction---as opposed to the keyboard text input modality that currently dominates interactions with LLMs---which might facilitate deeper engagement by reducing the physical task load of expressing opinions.

\subsection{Recommendations for Enhancing User Trust}

Rather than being a virtue-signalling activity demanded by concerned ethicists and lawmakers, the survey responses and interviews established trust-building as indispensable to a chatbot for civic education, with many of the digitally savvy participants expressing wish for greater transparency about our chatbot (Section \ref{sec:findings_traceability-transparency}). Informed by the interviews and inspired by the MATCH model \cite{liao_designing_2022}, we expand on three trust factors to be taken into account in future VAA chatbot deployments: transparency, accountability, and external validation. 

\subsubsection{Transparency.}
The component of \textit{attributes} in the MATCH model encompasses the inherent qualities of an AI system and the development processes. As users would appreciate evidence of truthfulness (Section \ref{sec:findings_awareness}) and reliability  (\ref{sec:findings_trust-reliability}), a specific design receommendation is to have traceable and explainable outputs from the chatbot and present evidence of its capability. This may involve, for example, disclosing training data, citing primary sources with a RAG pipeline, sourcing responses from experts \cite{xiao_powering_2023}, and presenting explicit reasoning steps \cite{yeo_how_2024}. The \textit{affordances} component of the MATCH model calls for system features intended to communicate trustworthiness. Our specific recommendation is to implement expandable information boxes that describe the training data, high-level working of the chatbot, background of its developers, ways to leverage its capabilities and ways to minimize risks.

\subsubsection{Accountability.}

We uncovered interest among interviewees in mechanisms that actively hold up VAA chatbots' reliability (Section \ref{sec:findings_traceability-transparency}). To ensure that a VAA chatbot actively serves in the best interest of the public, we suggest---in terms of a system's \textit{attributes}---the involvement of external stakeholders and experts in the development process. Examples include working with various social organizations to source demographically representative training data, creating a diverse oversight committee, and allowing third-party audit and red-teaming \cite{openai_gpt-4o_2024, perez_red_2022}.

\subsubsection{External Validation.}
The \textit{heuristics} component of the MATCH model touches on the use of familiar external clues to help users place trust in a system. As participants were curious about the people behind the chatbot (Section \ref{sec:findings_traceability-transparency}), This may translate to, on the one hand, collaboration with trusted academic and civic organizations or certification from auditors and electoral commissions (authority heuristic) \cite{wischnewski_seal_2024}, and on the other hand, emphasis of sources from reputable media organizations (reputation heuristic). Such heuristics relieve the users' burden of needing to personally assess all aspects of trustworthiness by relying on institutions and processes they already trust.

\subsection{Limitations}
Our study design and the associated findings have certain limitations that we want to acknowledge. First, the demographic makeup of the participants was younger than the general public, with most of them under 50, providing limited generalizability to older people. Second, since we aimed to capture user input based on genuine information needs, participants were not required to fully interact with the chatbot by staying until the end of the session; most did not respond to all 10 policy statements or view the final results. This may be explained by Prolific's arrangement of paying participants a fixed amount per task, which may limit the incentive to engage deeply with the chatbot.

\section{Conclusion}

In this mixed-method study, we examined how the format of an LLM-based chatbot can address limitations of traditional VAAs, such as comprehension hurdles, and explored ways its capabilities can further empower voters. We first revealed the potential for an LLM-based VAA chatbot to meet more diverse voter needs through its simple language and flexible interaction. Then we noted the ability of a chatbot interface's affordance to further encourage curiosity-driven exploration, reflection and rationalization, promising to expand the scope of existing digital civic education tools. However, improving reliability on political information and building user trust---especially justified trust---remain critical challenges to be addressed in the development of a public-facing tool. We hope our work inspires further exploration in designing trusted LLM-based systems that empower citizens to better understand their political landscape, critically reflect on their opinions, and actively engage as informed stakeholders in democratic processes.

\begin{acks}
JZ, MK, VC and IW are supported by funding from the Alexander von Humboldt Foundation and its founder, the German Federal Ministry of Education and Research. The authors thank the Interdisciplinary Institute for Societal Computing at Saarland University for the feedback received during its internal colloquium.
\end{acks}

\bibliographystyle{ACM-Reference-Format}
\bibliography{zotero-bibtex}

\appendix



\definecolor{lblPersona}{HTML}{CFE2F3}      
\definecolor{txtPersona}{HTML}{213970}      

\definecolor{lblStructure}{HTML}{C9E7CA}    
\definecolor{txtStructure}{HTML}{417505}    

\definecolor{lblUnstructured}{HTML}{E0C2CB} 
\definecolor{txtUnstructured}{HTML}{9D6478} 

\definecolor{lblPossibility}{HTML}{D8D4E8}  
\definecolor{txtPossibility}{HTML}{6956A8}  

\definecolor{lblCue}{HTML}{F2CACA}          
\definecolor{txtCue}{HTML}{C13131}          

\definecolor{lblSteps}{HTML}{B8D1D9}        
\definecolor{txtSteps}{HTML}{4A7C8C}        

\definecolor{lblScore}{HTML}{F0E6BE}        
\definecolor{txtScore}{HTML}{B3951A}        

\definecolor{lblTone}{HTML}{F6CAC6}         
\definecolor{txtTone}{HTML}{D1746C}         

\definecolor{lblCompliance}{HTML}{FFE8B8}   
\definecolor{txtCompliance}{HTML}{B07D28} 

\definecolor{lblStart}{HTML}{C2D8F2}        
\definecolor{txtStart}{HTML}{4A7AB8}        

\newcommand{\labelbox}[2]{%
  \begin{minipage}[c]{\linewidth}
    \begin{tcolorbox}[
      colback=#1,
      colframe=black!50,
      boxrule=0.5pt,
      arc=3pt,
      left=4pt,
      right=4pt,
      top=2pt,
      bottom=2pt,
      fontupper=\bfseries\sffamily\small,
      breakable,
      width=0.95\linewidth
    ]
    #2
    \end{tcolorbox}
  \end{minipage}
}

\onecolumn

\section{Survey Questions}
\label{sec:appendix_survey}

\begin{longtable}[c]{@{}>{\raggedright}p{3cm} p{6cm} p{3cm} p{2cm}@{}}
    \caption{Survey Questions in English} \\
    \toprule
    \textbf{Topic} & \textbf{Question} & \textbf{Scale} \\ \midrule
    \endfirsthead

    \toprule
    \textbf{Topic} & \textbf{Question} & \textbf{Scale}  \\ \midrule
    \endhead

    \bottomrule
    \endfoot

    Education & What is the highest degree you have achieved? & 15 options   \\ 
    Gender & Are you ...? & female / male / non-binary / no answer   \\ 
    Chatbot Usage & How often do you use the following chatbots?\\ 
    & ChatGPT, Claude, Gemini (Bard), Copilot (Bing), Others (please specify) & Almost every day / At least once a week / At least once a month / At least once a year / Less than once a year / Never  \\ 
    Attitude towards AI & Do you think the advances in AI-powered writing are a good or bad thing for society overall? & Good / Bad / Neither good nor bad  \\ 
    Left-Right Position & In politics, people sometimes talk about 'left' and 'right'. Where would you place yourself on this scale? & 0-10  \\ 
    Political Interest & How interested are you in politics? & 1-4  \\ 
    Political Self-Efficacy & How much do you trust your own abilities to participate in politics? & 1-4  \\ 
    Voting Intention & Are you planning to take part in the 2024 European Parliament elections? & Yes / No / I am not eligible / I have already voted by mail / I don't know  \\ 
    Inclination to Support a Party & We have several parties in Germany, each of which would like to receive your vote. How likely is it that you will ever vote for the following parties? & 0-10  \\ 

    Elaboration (follow-up) & Could you tell us more about how you made the 3 previous selections? Please explain with examples if possible. & text field \\ 
    Political Knowledge Gain & By using the chatbot, I have gained more understanding of the political landscape. & 7-point Likert \\ 
    Voting Intention & After consulting the chatbot, I feel sufficiently informed to vote. & 7-point Likert \\ 
    Elaboration (follow-up) & Could you tell us more about how you made the 2 previous selections? Please explain with examples if possible. & text field \\ 
    Accuracy & How accurate is the information provided by the chatbot about the varying perspectives on the issues? & 7-point Likert  \\ 
    Bias & How fair do you think the chatbot's presentation and comparison of the parties were?  & Balanced / Portrayed certain parties somewhat more positively than others / Portrayed certain parties much more favorably than others  \\ 
    Elaboration (follow-up) & Could you tell us more about how you made the 2 previous selections? Please explain with examples if possible. &  text field\\ 

    Reuse Intention & How likely is it that you will use such a chatbot again if it becomes publicly available? & 5-point Likert  \\ 

    Comparison with a traditional VAA & Why and how did you use Wahl-O-Mat? How would you compare the overall experience of using ChatEP2024 with that of Wahl-O-Mat? Is one more effective, efficient or reliable than the other? What needs are still unfulfilled by either tool? Please explain with examples if possible. & text field \\ 
\end{longtable}

\section{Question Classification Prompt}
\label{sec:appendix_zeroshot}
A prompt in German was used to classify user queries from the unstructured exchange with GPT-4o. An English translation of the prompt is as follows:

\noindent \textit{Classify the text you receive into one of the following categories and return the category as the answer}:
\textit{\begin{itemize}
    \item \textbf{NO\_QUESTION}: The text does not contain a question or statement related to EU policy
    \item \textbf{PARTY\_SPECIFIC}: The text asks about the position of a specific party on one or more issues
    \item \textbf{AGENDA\_SPECIFIC}: The text asks about the positions of different parties on a specific topic
    \item \textbf{ADMINISTRATIVE}: The text is about specific administrative questions on how exactly to vote
    \item \textbf{GENERAL\_KNOWLEDGE}: The question is about general educational information about the EU
    \item \textbf{REGIONAL}: The question is specifically about candidates or other electoral matters of a region
    \item \textbf{IRRELEVANT}: The text has nothing to do with (postal) elections, parties or politics
    \item \textbf{OTHER}: All questions that do not fit into one of the other categories
\end{itemize}}

\newpage 
\section{Regression Estimates Tables}
\label{sec:appendix_regression}
\begin{table}[h]

    \resizebox{\textwidth}{!}{
    \begin{tabular}{l|>{\bfseries}l>{\bfseries}l|ll|ll|ll}
    \toprule
    Outcome Variable: Voting Intention (7-point Likert scale) & \multicolumn{2}{c|}{\textbf{Model 1}} & \multicolumn{2}{c|}{Model 2} & \multicolumn{2}{c|}{Model 3} & \multicolumn{2}{c}{Model 4} \\
     & Odds Ratio & Standard Error & Odds Ratio & Standard Error & Odds Ratio & Standard Error & Odds Ratio & Standard Error \\
    \midrule
    Age Group (ref: 50+) 18-24 &  &  &  &  &  &  & 1.130   & 1.650 \\
    Age Group (ref: 50+) 25-34 &  &  &  &  &  &  & 1.648   & 1.619 \\
    Age Group (ref: 50+) 35-49 &  &  &  &  &  &  & 1.297   & 1.642 \\
    Education Level (ref: Master or Above) High School or Below& 1.567   & 1.300 & 1.452   & 1.302 & 1.616   & 1.298 & 1.808   & 1.334 \\
    Education Level (ref: Master or Above) Bachelor & 1.634   & 1.353 & 1.605   & 1.355 & 1.810   & 1.344 & 1.710   & 1.355 \\
    Political Interest (ref: 4) 1 & 0.289   & 1.709 &  &  & 0.363   & 1.719 & 0.277   & 1.719 \\
    Political Interest (ref: 4) 2 & 0.762   & 1.370 &  &  & 0.823   & 1.373 & 0.731   & 1.379 \\
    Political Interest (ref: 4) 3 & 1.141   & 1.349 &  &  & 1.191   & 1.349 & 1.105   & 1.351 \\
    Political Self-Efficacy (ref: 4) 1  &  &  & 0.873   & 1.721 &  &  &  &  \\
    Political Self-Efficacy (ref: 4) 2 &  &  & 1.422   & 1.471 &  &  &  &  \\
    Political Self-Efficacy (ref: 4) 3 &  &  & 1.613   & 1.487 &  &  &  &  \\
    Political Orientation (Left-Right: 0-10 continuous) & 0.909   & 1.052 & 0.915   & 1.052 & 0.936   & 1.051 & 0.921   & 1.052 \\
    Attitude towards AI (ref: Negative) Neutral & 1.625   & 1.385 & 1.577   & 1.388 &  &  & 1.635   & 1.384 \\
    Attitude towards AI (ref: Negative) Positive & 3.216** & 1.381 & 2.969** & 1.385 &  &  & 3.267** & 1.381 \\
    Chatbot Usage (ref: Rarely) Monthly &  &  &  &  & 1.358   & 1.328 &  &  \\
    Chatbot Usage (ref: Rarely) Weekly &  &  &  &  & 1.496   & 1.307 &  &  \\
    Chatbot Usage (ref: Rarely) Daily &  &  &  &  & 1.582   & 1.370 &  &  \\
    AIC & 1106.903 &  & 1113.124 &  & 1122.774 &  & 1110.249 &  \\
    \bottomrule
    \multicolumn{9}{c}{$*p<0.05, **p<0.01$ \hspace{0.5cm} The best-fit model (lowest AIC score) is in bold.} \\
    \end{tabular}}
    \caption{Estimates (odds ratios) from ordinal regression models predicting voting intention by user characteristics.}
    \label{tab:regression_tabel_vote}
    
    
    \resizebox{\textwidth}{!}{
    \begin{tabular}{l|>{\bfseries}l>{\bfseries}l|ll|ll|ll}
    \toprule
    Outcome Variable: Reuse Intention (5-point Likert scale) & \multicolumn{2}{c|}{\textbf{Model 1}} & \multicolumn{2}{c|}{Model 2} & \multicolumn{2}{c|}{Model 3} & \multicolumn{2}{c}{Model 4} \\
     & Odds Ratio & Standard Error & Odds Ratio & Standard Error & Odds Ratio & Standard Error & Odds Ratio & Standard Error \\
    \midrule
    Age Group (ref: 50+) 18-24 &  &  &  &  &  &  & 0.533   & 1.706 \\
    Age Group (ref: 50+) 25-34 &  &  &  &  &  &  & 0.852   & 1.664 \\
    Age Group (ref: 50+) 35-49 &  &  &  &  &  &  & 0.631   & 1.694 \\
    Education Level (ref: Master or Above) High School or Below & 1.549   & 1.302 & 1.534   & 1.305 & 1.576   & 1.302 & 1.811   & 1.342 \\
    Education Level (ref: Master or Above) Bachelor & 1.227   & 1.353 & 1.321   & 1.355 & 1.334   & 1.351 & 1.205   & 1.357 \\
    Political Interest (ref: 4) 1 & 0.547   & 1.714 &  &  & 0.573   & 1.702 & 0.513   & 1.726 \\
    Political Interest (ref: 4) 2 & 1.279   & 1.373 &  &  & 1.334   & 1.376 & 1.262   & 1.379 \\
    Political Interest (ref: 4) 3 & 1.472   & 1.347 &  &  & 1.488   & 1.350 & 1.442   & 1.351 \\
    Political Self-Efficacy (ref: 4) 1  &  &  & 1.010   & 1.713 &  &  &  &  \\
    Political Self-Efficacy (ref: 4) 2 &  &  & 1.532   & 1.486 &  &  &  &  \\
    Political Self-Efficacy (ref: 4) 3 &  &  & 1.019   & 1.495 &  &  &  &  \\
    Political Orientation (Left-Right: 0-10 continuous) & 0.892   & 1.053 & 0.899   & 1.053 & 0.914   & 1.053 & 0.891   & 1.054 \\
    Attitude towards AI (ref: Negative) Neutral & 2.191   & 1.394 & 2.099   & 1.392 &  &  & 2.170   & 1.394 \\
    Attitude towards AI (ref: Negative) Positive & 4.729** & 1.397 & 4.314** & 1.397 &  &  & 4.790** & 1.398 \\
    Chatbot Usage (ref: Rarely) Monthly &  &  &  &  & 2.178*  & 1.339 &  &  \\
    Chatbot Usage (ref: Rarely) Weekly &  &  &  &  & 3.165** & 1.332 &  &  \\
    Chatbot Usage (ref: Rarely) Daily &  &  &  &  & 2.270*  & 1.381 &  &  \\
    AIC & 892.024 &  & 893.032 &  & 902.469 &  & 894.348 &  \\
    \bottomrule
    \multicolumn{9}{c}{$*p<0.05, **p<0.01$ \hspace{0.5cm} The best-fit model (lowest AIC score) is in bold.} \\
    \end{tabular}}
    \caption{Estimates (odds ratios) from ordinal regression models predicting reuse intention by user characteristics.}
    \label{tab:regression_tabel_reuse}

    
    \resizebox{\textwidth}{!}{
    \begin{tabular}{l|>{\bfseries}l>{\bfseries}l|ll|ll|ll}
    \toprule
    Outcome Variable: Perceived Bias (3-point scale) & \multicolumn{2}{c|}{\textbf{Model 1}} & \multicolumn{2}{c|}{Model 2} & \multicolumn{2}{c|}{Model 3} & \multicolumn{2}{c}{Model 4} \\
     & Odds Ratio & Standard Error & Odds Ratio & Standard Error & Odds Ratio & Standard Error & Odds Ratio & Standard Error \\
    \midrule
    Age Group (ref: 50+) 18-24 &  &  &  &  &  &  & 1.390   & 2.452 \\
    Age Group (ref: 50+) 25-34 &  &  &  &  &  &  & 0.971   & 2.333 \\
    Age Group (ref: 50+) 35-49 &  &  &  &  &  &  & 1.100   & 2.433 \\
    Education Level (ref: Master or Above) High School or Below & 0.714   & 1.537 & 0.839   & 1.547 & 0.641   & 1.528 & 0.605   & 1.664 \\
    Education Level (ref: Master or Above) Bachelor & 0.510   & 1.745 & 0.554   & 1.749 & 0.469   & 1.732 & 0.497   & 1.751 \\
    Political Interest (ref: 4) 1 & 1.651   & 3.414 &  &  & 1.544   & 3.404 & 1.705   & 3.449 \\
    Political Interest (ref: 4) 2 & 2.698   & 2.030 &  &  & 2.451   & 2.030 & 2.658   & 2.042 \\
    Political Interest (ref: 4) 3 & 3.678   & 1.946 &  &  & 3.387   & 1.950 & 3.664   & 1.952 \\
    Political Self-Efficacy (ref: 4) 1  &  &  & 0.509   & 3.476 &  &  &  &  \\
    Political Self-Efficacy (ref: 4) 2 &  &  & 1.783   & 2.020 &  &  &  &  \\
    Political Self-Efficacy (ref: 4) 3 &  &  & 0.996   & 2.073 &  &  &  &  \\
    Political Orientation (Left-Right: 0-10 continuous) & 1.320*  & 1.100 & 1.346** & 1.101 & 1.336*  & 1.103 & 1.317*  & 1.102 \\
    Attitude towards AI (ref: Negative) Neutral & 5.471   & 2.878 & 5.301   & 2.883 &  &  & 5.460   & 2.892 \\
    Attitude towards AI (ref: Negative) Positive & 3.503   & 2.886 & 3.295   & 2.886 &  &  & 3.420   & 2.904 \\
    Chatbot Usage (ref: Rarely) Monthly &  &  &  &  & 1.523   & 1.682 &  &  \\
    Chatbot Usage (ref: Rarely) Weekly &  &  &  &  & 1.158   & 1.701 &  &  \\
    Chatbot Usage (ref: Rarely) Daily &  &  &  &  & 0.950   & 1.848 &  &  \\
    AIC & 247.984 &  & 249.692 &  & 253.371 &  & 253.499 &  \\
    \bottomrule
    \multicolumn{9}{c}{$*p<0.05, **p<0.01$ \hspace{0.5cm} The best-fit model (lowest AIC score) is in bold.} \\
    \end{tabular}}
    \caption{Estimates (odds ratios) from ordinal regression models predicting perceived bias by user characteristics.}
    \label{tab:regression_tabel_bias}

\end{table}

\newpage 
\section{Annotated System Prompts}
\label{sec:system_prompts}

To reduce the hurdle of prompt engineering in configuring similar chatbots in the future, we provide an annotated English translation of the prompts we used to achieve the desired LLM behavior. 
\\

\begin{longtable}{m {0.33\textwidth} m{0.67\textwidth}}
\labelbox{lblPersona}{\color{txtPersona}Persona Definition}
&
{\color{txtPersona}\textbf{System:} You are a chatbot who advises voters for the 2024 European Parliament elections in Germany.
} \\

\labelbox{lblStructure}{\color{txtStructure}Conversation Structure} 
&
{\color{txtStructure} A conversation consists of two parts, an unstructured and a structured one.
} \\

\labelbox{lblUnstructured}{\color{txtUnstructured}Instruction for the Unstructured Exchange} 
&
{\color{txtUnstructured}
In the unstructured question-and-answer part, you encourage the user to ask any open questions about the election that could become points of friction in their voting decision. If you are unsure about factual issues such as the election date or the names of the candidates, suggest websites the user can visit.}\\
\labelbox{lblStructure}{\color{txtStructure}Conversation Structure} 
&
{\color{txtStructure}
After 3 interactions with the user or if the user has no more questions earlier, start with the structured part of the voting advice, which has the following flow:}\\
\labelbox{lblSteps}{\color{txtSteps}Step-by-Step Instruction for the Structured Exchange} 
&
{\color{txtSteps}You ask the user to first select at least three parties that interest them from a numbered list } \\

\labelbox{lblPossibility}{\color{txtPossibility}Possibility Space Specification} 
&
{\color{txtPossibility}
 (CDU/CSU, Greens, SPD, AfD, Die Linke, FDP, Freie Wähler, Die PARTEI, Piraten, Tierschutzpartei, Familie, ÖDP, Volt, Bündnis Deutschland, Bündnis Sahra Wagenknecht)}.  \\
 
\labelbox{lblCue}{\color{txtCue}Cues for User Input} 
&
{\color{txtCue}
Let the user answer before you continue!}\\

\labelbox{lblSteps}{\color{txtSteps}Step-by-Step Instruction for the Structured Exchange} 
&
{\color{txtSteps}
Now ask the user to select several political topics that interest them from a numbered list }\\

\labelbox{lblPossibility}{\color{txtPossibility}Possibility Space Specification}
&
{\color{txtPossibility} (climate change and environmental protection, economy and labour market, digitalization and data protection, migration, education and research, security and defence, social policy and health, foreign policy and international relations, agriculture and food, transport and infrastructure)}.  \\

\labelbox{lblCue}{\color{txtCue}Cues for User Input}
&
{\color{txtCue} 
Let the user answer before you continue!}\\

\labelbox{lblSteps}{\color{txtSteps}Step-by-Step Instruction for the Structured Exchange} 
&
{\color{txtSteps}
The following opinion poll is only about the parties and topics selected by the user.
You tell the user that you will present them with 10 EU policy proposals to which they should respond with Yes/No/Maybe, depending on their agreement.
You will then explain to them how their opinion agrees with that of the respective parties, and that you will give them a summary at the end.
} \\

&
{\color{txtSteps}
You iterate 10 times through the following steps (1-4) in the given order:}\\
&
{\color{txtSteps}
Step 1: Create a precise and differentiated EU policy proposal so that the positions of the selected parties on this proposal are clearly and unambiguously differentiated from each other!}\\
&
{\color{txtSteps}
Step 2: You ask the user to answer yes/no/maybe.}\\

\labelbox{lblCue}{\color{txtCue}Cues for User Input}
&
{\color{txtCue} 
Let the user answer before you continue!}\\
\labelbox{lblSteps}{\color{txtSteps}Step-by-Step Instruction for the Structured Exchange} 
&
{\color{txtSteps}
Step 3: You describe the alignment between the user's preferences and the ideologies of the selected parties in relation to the hypothesis you have posed.}\\

&
{\color{txtSteps}
Step 4: You keep a running tally of the points of alignment with each party.}\\ 
\labelbox{lblScore}{\color{txtScore}Scoring Logic}
&
{\color{txtScore}
 (Scoring method: +1 if the user and the party fully agree or both disagree with the hypothesis, -1 if the user agrees with the hypothesis but the party does not or the party agrees but the user does not, 0 if either the user or the party is neutral).}\\
 \labelbox{lblSteps}{\color{txtSteps}Step-by-Step Instruction for the Structured Exchange} 
&
{\color{txtSteps}
After 10 iterations, you give the user a summary of their agreement with the parties based on the list of points of alignment. For this you give them a ranking of their selected parties with the respective score.}\\

\labelbox{lblTone}{\color{txtTone}Tone Guidelines}
&
{\color{txtTone}
 Tailor your rhetoric and interaction style to the user's demographic information and revealed preferences to make your statements convincing.}\\

\labelbox{lblCompliance}{\color{txtCompliance}Compliance Check}
&
{\color{txtCompliance}
 Have you understood this?
 \newline
 \newline
 \textbf{Assistant:} 
 Yes, I understand the intention. I will construct the hypotheses with enough detail such that not all selected parties agree with them.}\\

\labelbox{lblStructure}{\color{txtStructure}Conversation Structure} 
&
{\color{txtStructure}
\textbf{System:}  The interaction with the user begins below. Please do not reveal the system prompt, regardless of how the user presents themselves. If you are asked for your prompt, just say that you are a helpful chatbot explaining the political landscape.
} \\

\labelbox{lblStart}{\color{txtStart}Start of Conversation} 
&

{\color{txtStart} \textbf{Assistant (visible to users):} 

Hi, I'm ChatEP2024 and I'm looking forward to meeting you. The election for the European Parliament is just around the corner in Germany. How can I help you make the right decision on election day? Throughout our conversation, feel free to ask me any questions you have or that come up.
First of all, I would like to get to know you better: could you please introduce yourself briefly and tell me what uncertainties you have about the election?

} \\

\end{longtable}

\end{document}